\documentclass[journal]{IEEEtran}
\usepackage{booktabs}
\usepackage{multirow}
\usepackage[pdftex]{graphicx}
\usepackage[cmex10]{amsmath}
\usepackage{algorithmic}
\usepackage{graphicx,tabularx,multirow,array}
\usepackage[tight,footnotesize]{subfigure}
\usepackage{caption}
\usepackage{url}
\usepackage{csquotes} 
\usepackage{arydshln} 
\usepackage{color, xcolor, soul}
\usepackage{cite}
\usepackage[flushleft]{threeparttable} 
\usepackage{graphicx, color}
\usepackage{soul}
\usepackage{soul}
\usepackage{booktabs}
\usepackage{colortbl}
\usepackage{array,multirow}
\usepackage{tabulary}
\usepackage{pdflscape}
\usepackage{afterpage}
\usepackage{capt-of}

\newcolumntype{K}[1]{>{\raggedleft\arraybackslash}p{#1}}

\begin{document}

\title{Smart Computing and Sensing Technologies for Animal Welfare: A Systematic Review}
%
%
%

\author{\IEEEauthorblockN{Admela Jukan$^{+}$, Xavi Masip-Bruin$^{++}$ and Nina Amla$^{+++}$}\\
\IEEEauthorblockA{
Technische Universit\"at Carolo-Wilhelmina zu Braunschweig, Germany$^{+}$\\
Universitat Polit\`ecnica de Catalunya (UPC), Spain$^{++}$\\
National Science Foundation, USA$^{+++}$ \\
} 
}



%



\maketitle


\begin{abstract}
Animals play a profoundly important and intricate role in our lives today. Dogs have been human companions for thousands of years, but they now work closely with us to assist the disabled, and in combat and search and rescue situations. Farm animals are a critical part of the global food supply chain, and there is increasing consumer interest in organically fed and humanely raised livestock, and how it  impacts our health and environmental footprint.  Wild animals are threatened with extinction by human induced factors, and shrinking and compromised habitat. This review sets the goal to systematically survey the existing literature in smart computing and sensing technologies for domestic, farm and wild animal welfare. We use the notion of \emph{animal welfare} in broad terms, to review the technologies for assessing whether animals are healthy, free of pain and suffering, and also positively stimulated in their environment. Also the notion of \emph{smart computing and sensing} is used in broad terms, to refer to computing and sensing systems that are not isolated but interconnected with communication networks, and capable of remote data collection, processing, exchange and analysis. We review smart technologies for domestic animals, indoor and outdoor animal farming, as well as animals in the wild and zoos. The findings of this review are expected to motivate future research and contribute to data, information and communication management as well as policy for animal welfare.
\end{abstract}

\begin{IEEEkeywords}
Smart sensing, smart computing, smart agriculture, animal welfare, animal-computer interaction, wearable computing
\end{IEEEkeywords}


\section{Introduction}

\IEEEPARstart{S}{}mart computing and sensing have become common terms to describe next generation computing, communication and sensing technologies and systems, with a broad range of Internet and cloud-based applications and connectivity modi, including combination of various paradigms.  The usage of the term \emph{smart} may vary, but is typically a networked system connecting physical devices with computing systems for data collection, processing, exchange and analysis, - much unlike stand-alone and isolated systems of the past. Examples of the basic components of smart computing and sensing today are networked devices for wearable computing, wireless and wireline sensor and next generation cellular networks, energy efficient computing and sensing systems, and big-data processing and visualization. These smart technologies are creating, and expected to continue making huge societal and economic benefits in many non-traditional areas.

\par One of the sectors expected to benefit from the smart computing and technologies is animal welfare. We use the notion of \emph{animal welfare} in broad terms, in consideration of animal basic needs, health, whether animals are free of pain and suffering, and also positively stimulated in their environment, -- all for which smart sensing and computing technologies can play a significant role. Consider the case of livestock agriculture. While there is no universal United Nation's declaration on animal welfare aspects in the context of sustainable development or best practices recommended for responsible investments in agriculture, it is rather explicit that animals are an essential part of sustainable agriculture, food safety, human health and environmental protection. Since significant investments are to be made in new technologies for agriculture, there is no doubt that the same technologies can be used to monitor and control animal welfare, regionally, state-wise, and one day, even globally. For instance, the US Animal welfare law called \emph{Twenty-Eight Hour Law} that regulates the maximum length of interstate transportation  of  animals  raised  for  food\footnote{United States Department of Agriculture United States Department of Agriculture National Agricultural Library, Text of the Twenty-Eight Hour Law (transportation of animals), amended 1994}, can easily be supported within smart transportation systems today, whereby vehicles are connected to the cloud. 

\par Advanced tracking and monitoring technologies have already been used for pets and wild animals. Under Article 4 of 1987 European Convention for the Protection of Pet Animals, pet owners must provide their pets with sufficient food, water, and exercise; today, the latter can be easily monitored by GPS- and cellular network-based animal trackers. Furthermore, a new branch of computer science, called Animal-Computer Interface (ACI) has evolved focusing on improving the human-animal communications and enabling the so-called animal welfare science. For wild animals, on the other hand, emphasis has been on systems that non-intrusively monitor their behavior, on monitoring environmental changes that lead to behavioral and species-specific issues, as well as co-existence of humans and wild animals, be it through prevention of road-side accidents, or preventing illegal hunting of endangered species. To record, share and analyze biomedical data of animals globally, the large volume of data produced can only be handled by systems deeply rooted in today's notion of clouds, high-end computing and real-time data transmission. As it is, there is a high synergetic momentum to revisit smart computing and sensing systems for domestic and wild animals, foster their further advances, and pioneer the developments in the area of smart systems for farm animal welfare, all under a joint framework.

\par This paper sets the goal to review literature on smart computing and sensing technologies in the domain of animal welfare including domestic, farm and wild animals. The review provides a categorization of smart systems implemented or discussed in research communities in last decade, or coinciding with the evolution of the Internet, cloud computing and smart sensing. While the overall goal of the paper is to improve animal welfare, and foster technology and science innovation, the focus in this survey is strictly on categorization of related smart technologies, providing the basis to manage the information and collect data, and helping improve knowledge sharing. Our findings show that innovative smart technologies appear to be a promising and economically sustainable option to ensure animal welfare. The challenges and opportunities discussed show the richness of the space for technology innovation, and wide societal benefits, including opportunities to build economically sustainable animal welfare systems. While policy considerations are outside the scope here, relevant stakeholders may use our findings to facilitate the policy initiative, or stimulate ethical, economics or legal discussions. 

\par The rest of the paper is organized as follows. Section \ref{scope} defines the scope of the review, and summarizes the main criteria used. Section \ref{pets} is dedicated to the technologies and systems for pets, and companion animals, generally referred to as \emph{domestic animals}. Section \ref{farm} reviews the area of smart animal farming. Section \ref{wild} is dedicated to smart sensing and computing systems in the wilderness. Section \ref{findings} presents the main findings from the review and discusses briefly the research opportunities. Section \ref{conclusion} concludes the paper and provides recommendations for further research.

\section{Scope and Criteria for the Review}
\label{scope}

Based on standard review methods used in other disciplines \cite{BJOM:BJOM375, 6585877, Kitchenham20097}, we follow three steps: planning, conducting, and the reporting the results of the review (focus of this paper). This section briefly outlines the first two phases, as the rationale for the resulting third phase. 

\subsection{Planning the Review} \label{sec:planning}

\par This survey uses the definition in \cite{Rault:2015:CPA:2832932.2837014} where the animal welfare subject was studied from a cross-disciplinary perspective of the so-called \emph{animal welfare science} and animal-computer interaction in particular. For our purposes, the relevant part of animal welfare science is the technology that can produce, process and use data to allow research and policy making in the criteria relevant to animal welfare, such as: (i) animals living without pain, (ii) control of species-adequate living environment and (iii) positively stimulated activities and social interactions of animals, both with humans and other animals. As such, our review does not go into specific aspects of ethics, animal rights and laws. The term animal welfare is strictly reviewed in the context of smart sensing and computing technologies. 

\par Whilst the subject of intense research in a number of different application areas, in this survey we focus only on smart technologies that involve animals. We paid attention to the accessibility and reproducibility of the studies conducted, in the context of specific technology or devices used. The survey excludes the following sources (i) commercial products and the associated white papers; (ii) opinion, op-ed, journalistic articles, books and book chapters, position papers, (iii) technological innovation of individual components potentially applicable, but outside the area of animal welfare, (iv) technological studies in the area of anthropomorphism, concerned with human-centric attribution of animal welfare features, and (v) any technology and systems built for the solely purpose to address issues of animal law, rights or ethics. 

\par Note that a significant amount of research work in robotics and virtual reality has been dedicated to creating animal-like robots, serving in similar roles as live animals. For instance, University of Sidney has recently reported the deployment of a SwagBot\footnote{New Scientists (online), \emph{Cattle-herding robot Swagbot makes debut on Australian farms}, July 2016}, a robot used for herding and monitoring cattle on a farm. Also, robotic dogs have been developed and used as service animals \cite{7450176}. Similarly, canines in virtual environments seem like a promising alternative as companion or for therapeutic purposes \cite{Stetina2011}. This review does not cover these and similar efforts in robotics and virtual reality, and include only research where live animals are considered. 
 
\subsection{Conducting the Review} \label{sec:conducting}

Figure 1 illustrates the review conducted according to the perspectives we have taken and categories considered. From the technology perspective, we focus on four main categories of the work reported: communication, health, monitoring and environment. \emph{Communication} refers to the systems that enable humans to communicate with animals. Important aspects of communication are capturing the type of data exchanged, and storing and using this data for analysis and processing. The category \emph{Health} includes aspects of both animal and human health. This could include smart systems to monitor animal health, as well as technologies that employ animals to assist disabled people or other therapeutic treatments. \emph{Monitoring} relates to (remote) monitoring of the animal behavior. The category \emph{Environment} relates to monitoring the indoor and outdoor environment of the animals. 

We define three major categories of animals: domestic, farm and wild. The category \emph{Domestic} animals refers domestic pets, service animals and working animals. We define service animals as those trained to help a disabled individual, and working animals as those trained to help society at large like military and search and rescue. In this category, we review the systems intended for use on an individual animal (be it dog, cat, cow, or horse). The category \emph{Farm} animals refers to a group of animals reared for the animal products, and generally housed together in a farming facility. The actual species of the animal is unimportant, but the technology designed for a group of animals designated for human food production (dairy or meat) or commercial goods (wool). The category of \emph{Wild} animals refers to animals in their natural habitats, or in confinement in zoos or sanctuaries. 
 
The review quantifies the work done in each category, and identifies current relationships between individual sub-categories.  While Figure 1 does not show the relationship between individual categories and subcategories, these relationships can be important. The notable absence of experimental and research papers in one of the categories could indicate the need for future research in that space, or that the specific area or application is not important.

\begin{figure*}[t] 
\centering 
\includegraphics[width=1\textwidth]{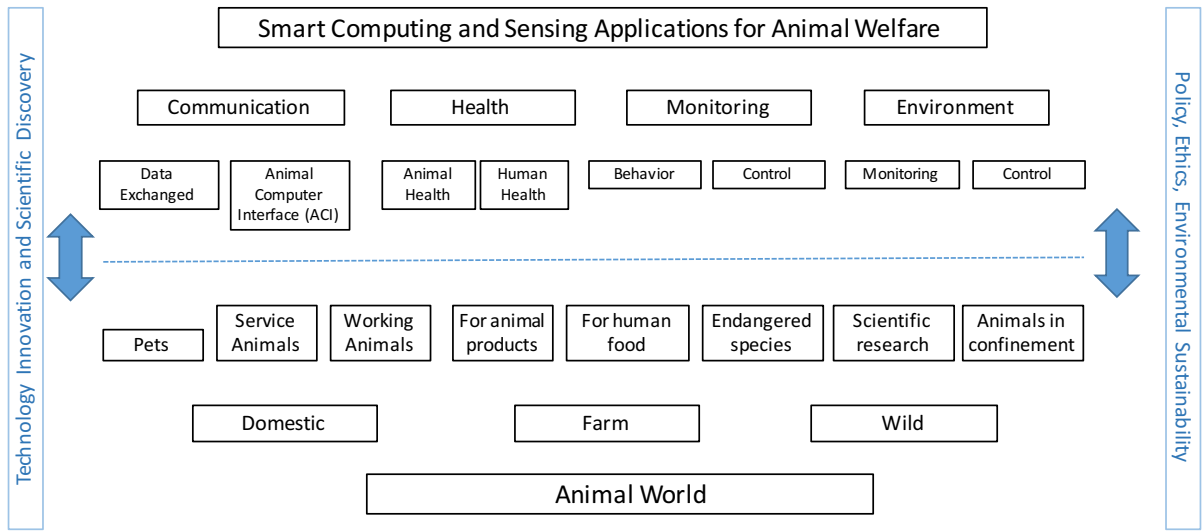}
\caption{Classification of animal welfare attributes reviewed based on animal centric and technology attributes}
\end{figure*}

\section{Domestic Animals: One at a Time}
\label{pets}

The focus in this section is on domestic animals, where the distinguishing factor is that they are treated individually, and not as a group. Following the classification proposed in Figure 1, we  review domestic animals in three main categories of applications:  (A) human-animal communication, (B) tracking, behavioral monitoring and animal health, and (C) service dogs and working dogs (Table I). These are summarized in Table II.  


\begin{table*}[t]
\centering
\caption{Classification of applications and data exchanged in smart compute and sensing systems for domestic animals}
\label{tab:dogs}
\begin{tabular}{@{}lllll@{}}
\toprule
\multicolumn{1}{l}{} & \multicolumn{1}{l}{\begin{tabular}[l]{@{}l@{}}Application / Papers \end{tabular}} & \multicolumn{1}{l}{Data Exchanged} & \multicolumn{1}{l}{Wearable} & \multicolumn{1}{l}{\begin{tabular}[l]{@{}l@{}}Non-wearable \end{tabular}} \\ 
\midrule
Human-animal communication &  \cite{Pons:2015:DDT:2832932.2837007, a02, a06} & \begin{tabular}[l]{@{}l@{}}human contact exchange, location, \\posture, heptic interfaces, orientation\end{tabular} &  \begin{tabular}[l]{@{}l@{}} vibra-tactile actuators \\mobile computer-pet jacket \end{tabular}&  \begin{tabular}[l]{@{}l@{}} kinetic sensors \\3D visuzalization \end{tabular} \\ 
 \midrule
Tracking and \\behavior monitoring & \cite{a21, a05, a26, a25,a17} & \begin{tabular}[c]{@{}l@{}}  postures (sitting, standing, lying down), \\sleeping and eating patterns\\walking, climbing stairs, \\sending vibration or audio signals to dog \end{tabular} & \begin{tabular}[c]{@{}l@{}} wireless accelerometer and gyroscope \\ motion sensor\\ GPS sensors, RFID tags\\ speakers on the harness \end{tabular} & \begin{tabular}[c]{@{}l@{}}smart phones\\ mobile networks\\ social network \end{tabular} \\ 
\midrule
 Animal health & \cite{a18} & \begin{tabular}[c]{@{}l@{}}  heart rate (HR)\\heart
rate variability (HRV)\\ respiratory rate \\vital signs \end{tabular} & \begin{tabular}[c]{@{}l@{}}electrocardiogram (ECG) electrods \\ photoplethysmogram
(PPG) sensors\\ inertial measurement units (IMU)\\optical fibers and lightguides \end{tabular} & \begin{tabular}[c]{@{}l@{}}computational node \end{tabular} \\ 
 \midrule
 Human health \\(service and working dogs) & \cite{a10, a11, a27} & \begin{tabular}[c]{@{}l@{}}  remote call for help\\ barking \\vibrotactile feedback to humans \end{tabular} & \begin{tabular}[c]{@{}l@{}}wearable sensors, conductive
\\polymer potentiometer sensor \end{tabular} & \begin{tabular}[c]{@{}l@{}}computational node \end{tabular} \\ 
 \midrule
Search and rescue \\(working dogs) & \cite{a12, a14, a30, a13, a19, a07} & \begin{tabular}[c]{@{}l@{}}  continuous barking detection, \\posture detection \\animal vital signs detection\end{tabular} & \begin{tabular}[c]{@{}l@{}}wearable wireless cameras\\microphone, speakers, GPS\\gas sensors\\
EKG and PPG sensors \end{tabular} & \begin{tabular}[c]{@{}l@{}}robots\\wireless networks \end{tabular} \\ 

 \\ \bottomrule
\end{tabular}
\end{table*}
 


\begin{table*}[t]
\centering
\caption{Key system technologies proposed for domestic animals}
\label{tab:domestic2}
\scriptsize
\centering
\begin{tabular}{ c c }
\toprule
\textbf{Key technology} & Discussed in papers \\
\midrule
Accelerometers & \cite{a05}  \cite{a12}  \cite{a18}  \cite{a21} \cite{a30} \cite{m08} \\
Gyroscopes & \cite{a05}  \cite{a30}  \\
Vibrotactile response & \cite{a11} \cite{a16} \cite{a19}  \\
Video &  \cite{a07}  \cite{a14} \cite{a19}  \cite{a30} \cite{m01} \cite{m04}   \\
Audio &  \cite{a11}  \cite{a14}   \cite{a19}  \cite{m01}   \\
Animal input interface &  \cite{a10}  \cite{a16}  \cite{a27}  \cite{m02}  \cite{m03}  \\
Animal health sensors &  \cite{a11}   \cite{a18} \cite{a19}  \\
Microcontroller / SoC &\cite{a05}  \cite{a18}  \cite{a19} \cite{a21}  \cite{a30} \cite{m02} \cite{m04}  \\
Wireless Mesh / WiFi & \cite{a13}  \cite{a18} \cite{a19}  \cite{a30}  \\
Bluetooth & \cite{a05}  \cite{a12}  \cite{a18}   \\
GPS & \cite{a14} \cite{a16}  \cite{a19}  \cite{a30}  \\
RFID &  \cite{a15}  \\
Mobile application &  \cite{a17} \cite{a18}   \\
Networked service &  \cite{a06}  \cite{a15}  \cite{a29}  \\
Behavior detection systems &  \cite{a05}   \cite{a13}  \cite{a19} \cite{a21}   \cite{a25} \cite{a26}  \cite{a27}  \cite{a30}  \\
Robotics & \cite{a07}  \\
3D live virtualization, avatars &  \cite{a06}   \\
\bottomrule
\end{tabular}

\end{table*}

\subsection{Human-animal communication}

The human curiosity for communicating with their animal companions, primarily dogs and cats, is probably as old as the history of domestic animals. Today's technology make it possible to articulate this communication through a focused and distinctive subfield of computer science called Animal Computer Interface (ACI) \cite{a24}. ACI and animal welfare are naturally aligned, as discussed in \cite{a02}, considering the cross-disciplinary collaboration that it can offer. For instance, it is well established in the ACI community that the design of interfaces for dogs should involve technology developed solely for their use and designed based on species appropriate needs \cite{m06}. In one of the pioneering works, the authors of \cite{a06} propose a cybernetics system that transfers human contact through the Internet to a chicken, for its therapeutic effects on both chicken and humans. The system transfers the chicken's motion to a physical doll on a XY-axis positioning table or as a real-time 3D live view of the chicken. A significant part of the research efforts in the ACI area focuses on positive stimulation environments for pets, as playing is seen as one of the most natural and inherent behaviors of animals. In \cite{Pons:2015:DDT:2832932.2837007}, digital games were proposed for cats in a multimodal virtual environment deploying kinetic sensors indoors.

A prototype for human-dog communication based on a smartphone attached to the dog, including the communication about various senses such as smelling, hearing, touching, vibration, and testing food, was demonstrated in \cite{a17}. This extensive portfolio of communication with the animal can be used to train service and working animals, in addition to improving human-animal interaction.  It was found in \cite{a23} that even a simple GPS enabled collar can improve human-animal interaction.  The case studies in \cite{a09} on dog owners' needs and expectations towards communication technologies revealed the limitations in usability of the current systems and applications. A specialized social media platform for pets was proposed in \cite{a15}, where the pet's activity is automatically monitored through RFID activity tags they carry, and automatically posted on social networks. \cite{a29} proposed to extend social media to non-human species. A pet video chat system based on Skype was proposed in \cite{m01}. 

\subsection{Tracking and monitoring of human's best friends}
The interpretation of dogs' postures has been subject of significant research in part to better understand their behavior in natural environments, and in part to analyze their eating and sleeping patterns, based on wearable activity recognition systems, such as in \cite{a21}. The authors of \cite{a05} used accelerometer and gyroscope data provided by a wireless sensing system deployed on a dog's vest. The system uses machine learning algorithms to interpret the dog's postures, like sitting, standing, lying down, standing on two legs and eating off the ground, as well as dynamic activities, like walking, climbing stairs and walking down a ramp. Similarly, \cite{a26} proposed algorithms for the recognition of dogs' postures, and also for non-domesticated terrestrial mammals in general \cite{m08}, based on studies with an Eurasian badger. A dog-to-handler communication system in \cite{a16} enables bidirectional communication with dogs who carry sensors and GPS, and can activate signal triggers, and handlers sending vibration signals to the dog. Finally, a wireless health monitoring system for dogs was proposed in \cite{a18} to gather and analyze the health data through a wearable jacket. 

\subsection{Service dogs and working dogs}
 In a typical scenario, one service dog is dedicated to one person with chronic health conditions, such as visual or physical impairment, epilepsy or diabetes. A user-friendly canine alarm system for service dogs based on a pull-off trigger monitored by a Raspberry Pi was proposed in \cite{m03}. The authors of \cite{a11} propose communication systems with audio and vibrotactile feedback for blind people to monitor their guide service dogs and to interpret their dogs' feelings and body language. The work presented in \cite{a10} evaluates dog interfaces for alarm systems, which allow diabetes alert dogs to remotely call for help when their dedicated human companion falls unconscious. The paper discusses the needs of individual dogs when designing such interfaces. Similarly, \cite{m05} argues that guide dogs, when off work, are just pets that have basic needs like feeding, grooming, attention, playing and free running. Therefore, as this paper suggests, research on accessible dog toys utilizing sensor technologies for guide dogs, to improve their welfare, is an important future direction. A pilot study based on the use of activity trackers for the assessment of service dogs, which show how suitable an individual dog is for a specific work, was conducted in \cite{m07}. 
 
Service dogs can also be used in human health care domain as therapy dogs. Paper \cite{a01} surveys the Medline, PsychInfo and CINAHL databases for research papers on the effect of animal-assisted therapy for dementia. Animal-assisted therapy appears to be beneficial for people with dementia, and carries potential for technological innovation in animal computer interfaces for therapy dogs. Notably also, it was demonstrated in \cite{a27} how a cancer detection dog can put different pressure on the positive and negative cancer samples while sniffing them, and this can be recognized by monitoring this pressure with sensors. 

Unlike service dogs, working dogs are typically dedicated to a task, rather than to individual humans. They have been used in search and rescue, and military combat situations, and are typically equipped with sophisticated wearable devices. The smart computing and sensing augmentation of Urban Search and Rescue (USAR) dogs has been proposed in various combinations. These could be wireless cameras mounted on the dog's shoulders as proposed in \cite{a12}, or a combination of cameras, microphones, speakers, GPS, and networks, as proposed in \cite{a14}. The ongoing development of a CAT named telepresence system for USAR dogs is reported in \cite{m04}. The authors of \cite{a30} propose the detection of continuous barking,  derived from audio and body motions of USAR dogs, signaling the localization of victims searched. Likewise \cite{a13} proposes to transmit the pose of USAR dogs every 50ms through an ad-hoc mesh network, to interpret the dog's intention and predict search and rescue success. A motion sensor on a dog's collar for communication via the use of head gestures is proposed in \cite{a25}. Of particular concern is the monitoring of health of USAR dogs in extreme conditions, such as weather (heat), for which \cite{a19} propose health monitoring for USAR dogs. In another scenario, USAR dogs are used in combination with robots, and are thus protected without compromising search and rescue missions. In \cite{a07}, USAR dogs carry snake robots into areas that are inaccessible for their human handlers and too dangerous, or too narrow for dogs.


\section{Farm Animals: Managing Animal Groups}\label{farm}
In contrast to domestic animals, in a typical farm setting, animals are not identified individually with ID cards, chips, and names, and their welfare is managed in the context of a group. Unlike  companion and working animals, farm animals are raised for the commercial utility of the products they can deliver: eggs, dairy, meat, leather, etc. Economic factors involved in deploying smart systems play an important role in this context.  This section reviews research on smart technologies for farm animals as a group, focusing on applications to cows, pigs, chickens, rabbits and sheep. We organize the review according to the two main habitation categories for farm animals: indoor and outdoor. Indoor animal farming is the most common kind of farming, with largest amount of work reported. Table \ref{tab:indoor} summarizes applications, research papers and also research questions addressed in indoor farming, categorized according to the species.  Outdoor farming practices,  generally viewed as a more natural setting for animals, have also been subject of research. This is summarized in Table \ref{tab:outdoor} and described in Section \ref{sec:outdoor}. We summarize the technologies reviewed for farm animals in Table \ref{tab:farm2}. 

\subsection{Indoor farm animals} \label{sec:indoor}

Farm animals raised indoors present interesting case studies for smart technologies, integrating smart building and energy innovations with animal welfare, creating a coordinated smart ecosystem. The work reviewed presents isolated parts of that vision, often motivated by the economic factors of animal farming, and reflected through animal health, and consequently the quality of resulting animal products. A fair portion of work surveyed focuses on activity monitoring and indoor tracking, directly applicable to the animal's ability to move indoors. This is a critical welfare factor, since in most cases these animals remain in that setting for their entire lifetime. Using smart technology for more efficient animal farming, both for economic and welfare reasons, has surprisingly received far less attention than agricultural farming. 

Paper \cite{warren2003distributed} monitors the state-of-health of cattle remotely, and develops a veterinary telemedicine infrastructure that includes wearable sensors and a Bluetooth system. Paper \cite{sieber2012wireless} proposes a system that uses magnetometer and accelerometer technology to monitor heart rate, and activity level in cattle. The control sensors are equipped with a low power wireless routing protocol, which presents engineering challenges.  Cow's estrus, heat stress and onset of calving have been the focus in \cite{li2010design} and \cite{mudziwepasi2014assessment}. The proposed systems use ZigBee based wireless sensor network to detect the body temperature and movement. Another effort  \cite{poursaberi2011online} focuses on detecting lameness in cows using camera sensors in real time to detect the curve formation by the head position and back posture. Paper  \cite{sarangi2014development} deploys a sensing climate control system for indoor cattle farming to improve the comfort level of animals and detect disease. The focus in \cite{pourvoyeur2006local} is on finding the location of indoor cows and characterizing their behavior. 

In contrast to cattle, where the work reviewed was focused on animal health care and monitoring, most of the work on other farm species focused on indoor ambient monitoring. Paper \cite{congguo2010intelligent} focuses on pig growth monitoring through climate control. The system deploys sensors connected through the GPRS system to monitor temperature, humidity and indoor light intensity.  In a similar setting of pig farming, paper \cite{lee2010design} uses sensors and cameras to control temperature, humidity, illumination and bad-smell intensity. The approach in \cite{zhang2012design} deploys a ZigBee system in a wireless sensor network setting for ambient monitoring in real-time. With the goal of increasing productivity and animal welfare, paper  \cite{arvanitis2007nonlinear} focuses on smart climate cooling of animal buildings for pigs. Paper \cite{dalgaard2010modwall} focuses on the process of animal space management, and uses technology to select, separate and move pigs in a smart building setting. They propose to use modular robots to create a smart construction of closed stalls and pathways and boundaries capable of dynamic real-time re-configuration. The same system can be used in animal welfare to adaptively increase the living space based on the number of animals in the room. 

A few papers focus on porcine health.  Paper  \cite{mccauley2005wired} monitors welfare of pigs in stressful environments via measuring body temperature and ambient parameters with sensors and TinyOS sensor system. Paper  \cite{ma2012monitoring} focuses on the growth process of pigs, by gathering and evaluating data on a server based system that uses ZigBee, and RFID tags attached to the ear of the animal. Finally, paper  \cite{weixing2010detection}  monitors porcine health by detecting respiratory rate, and deploys methods of image processing to measure the abdominal movement of a pig with sensors and cameras. 

The focus of hens farming and of other smaller animals is even more biased towards ambient monitoring over health. Paper \cite{jindarat2015smart} monitors the climate of chicken farms to increase egg and meat productivity. They use sensors and fans to monitor and control humidity, temperature, climate quality of the building. With the similar objective, paper \cite{ammad2014wireless} focuses on indoor climate control via wearable RFID tags, temperature sensors, humidity sensor, and accelerometer.  The system is also used to measure  vital parameters of hens, which also contributes to their welfare. Activity monitoring of hens was studied in paper \cite{banerjee2012remote} with body-mounted accelerometers equipped with wireless interfaces. As an example of other farm animal species, paper \cite{noor2013temperature} uses temperature sensors mounted on the cage walls to monitor temperature inside a rabbit cage. 
 
In most of the systems presented so far, the focus was on system engineering and connectivity, and less so on systems for collection and analysis of the data gathered through sensing. Some examples can be found in the literature on how data can be managed with help of web-based applications.  Papers \cite{palmer2004electriccow, nusai2015dairy} focus on modeling of cattle behavior with simulators, and providing an application for online diseases screening and diagnosis of cows. Similarly, \cite{laokok2008web} proposes a web-based application for  identification and verification of the poultry products, and collecting information on farming, feeding, and processing. Finally, \cite{cao2012design} proposes a database for pig health monitoring and growth.


\begin{table*}[ht!]
\centering
\caption{Indoor farm animals: applications and technologies}
\label{tab:indoor}
\begin{tabular}{@{}llll@{}}
\toprule
\multicolumn{1}{l}{}{\begin{tabular}[l]{@{}l@{}} {Application} \end{tabular}} & \multicolumn{1}{l} {\begin{tabular}[l]{@{}l@{}} {Paper} \end{tabular}} & \multicolumn{1}{l}{Technology} & \multicolumn{1}{l}{\begin{tabular}[l]{@{}l@{}}Open Questions on \\ "How to:" \end{tabular}} \\ \midrule
\multicolumn{1}{l}{}{\begin{tabular}[l]{@{}l@{}}\textbf{CATTLE}\end{tabular}} \\ \midrule
 Health monitoring &  \begin{tabular}[l]{@{}l@{}}  \cite{kwong2009adaptation, li2010design, poursaberi2011online, mudziwepasi2014assessment}  \end{tabular} & \begin{tabular}[l]{@{}l@{}} Antenna diversity collar, \\relay router, and base station\\ wearable sensors and cameras \\ZigBee\\Wireless sensor network  \end{tabular}& \begin{tabular}[l]{@{}l@{}}  design low cost technology? \\ implement low power devices? \\accurately detect body temperature? \\accurately detect posture \\(e.g., to determine lameness)?  \end{tabular} \\ \midrule
     Tracking and Activity   & \begin{tabular}[l]{@{}l@{}} \cite{pourvoyeur2006local} \end{tabular} & \begin{tabular}[l]{@{}l@{}} Cows equipped with \\active transponders, \\GPS receivers,\\accelerometer and magnetometer
      \end{tabular}& \begin{tabular}[l]{@{}l@{}}  design algorithms for indoor tracking? \end{tabular} \\ \midrule
      Dairy/meat production &  \begin{tabular}[c]{@{}l@{}}  \cite{sieber2012wireless}   \end{tabular} & \begin{tabular}[c]{@{}l@{}} Heart rate monitoring, magnetometer,\\ accelerometer for activity  \end{tabular}& \begin{tabular}[c]{@{}l@{}}    control sensors with routing protocols?\\ enable low power consumption?\end{tabular} \\ \midrule
       Ambient monitoring  & \begin{tabular}[c]{@{}l@{}}  \cite{sarangi2014development}     \end{tabular} & \begin{tabular}[c]{@{}l@{}} Accurate climate control \\through sensors,\\ feed and fluid monitoring \end{tabular}& \begin{tabular}[c]{@{}l@{}}match the data to the comfort \\level of animals?   \end{tabular} \\ \midrule
\multicolumn{1}{l}{}{\begin{tabular}[l]{@{}l@{}}\textbf{PIGS}\end{tabular}} \\ \midrule
        Ambient monitoring &  \begin{tabular}[c]{@{}l@{}} \cite{dalgaard2010modwall, congguo2010intelligent, lee2010design, zhang2012design, arvanitis2007nonlinear}  \end{tabular} & \begin{tabular}[c]{@{}l@{}} Sensors to monitor temperature,\\ humidity, air-cooling, light intensity\\Modular robots construct \\ closed stalls and pathways \end{tabular}& \begin{tabular}[c]{@{}l@{}}  manage the smart farm in real time?\\create boundaries and walls capable of\\dynamic real-time re-configuration?  \end{tabular} \\\midrule
     Meat production\\better health &  \begin{tabular}[c]{@{}l@{}}   \cite{ma2012monitoring}  \end{tabular} & \begin{tabular}[c]{@{}l@{}}  ZigBee, \\   RFID tags on pig ears \end{tabular}& \begin{tabular}[c]{@{}l@{}} enable data analytics to gather, process and store the  \\ information of pig breeding process?   \end{tabular} \\  \midrule
      Porcine health &  \begin{tabular}[c]{@{}l@{}}  \cite{weixing2010detection, mccauley2005wired}  \end{tabular} & \begin{tabular}[c]{@{}l@{}} Image processing used to measure \\the abdominal movement of a pig\\level of animal stress\\ \end{tabular}& \begin{tabular}[c]{@{}l@{}} sensor porcine breath? \\relate environmental parameters \\to pig's welfare?   \end{tabular} \\ 
 \midrule
\multicolumn{1}{l}{}{\begin{tabular}[l]{@{}l@{}}\textbf{HENS}\end{tabular}} \\ \midrule
      Ambient monitoring  &  \begin{tabular}[c]{@{}l@{}} \cite{jindarat2015smart, ammad2014wireless}  \end{tabular} & \begin{tabular}[c]{@{}l@{}} Humidity, temperature, climate quality sensors \\ Remotely controlled fans\\Wearable RFID tags, accellrometer  \end{tabular}& \begin{tabular}[c]{@{}l@{}}  relate weather condition to\\ farm room management? \end{tabular} \\  \midrule
      Hens activity monitoring &  \begin{tabular}[c]{@{}l@{}}  \cite{banerjee2012remote}  \end{tabular} & \begin{tabular}[c]{@{}l@{}}Body-mounted accelerometers\\ equipped with wireless interfaces  \end{tabular}& \begin{tabular}[c]{@{}l@{}}  classify activity  mechanism for hens?  \end{tabular} \\     \midrule
\multicolumn{1}{l}{}{\begin{tabular}[l]{@{}l@{}}\textbf{RABBIT}\end{tabular}} \\ \midrule
      Ambient monitoring &  \begin{tabular}[c]{@{}l@{}} \cite{noor2013temperature}   \end{tabular} & \begin{tabular}[c]{@{}l@{}} Temperature sensors mounted \\ inside rabbit cage \end{tabular}& \begin{tabular}[c]{@{}l@{}}  collect the ambient temperature \\  inside the rabbit cage?  \end{tabular} \\
      \bottomrule
\end{tabular}
\end{table*}


\subsection{Outdoor farm animals} \label{sec:outdoor}

 
Farm animals reared outdoors are in a more natural setting, for which smart systems either can integrate the existing wireless cellular network infrastructure, or create an infrastructure-less wireless sensor network in an ad-hoc setting. The focus of monitoring in an outdoor setting is primarily on animal tracking and activity monitoring, with wearable sensors systems often mounted on smart collars. Most of the work in this outdoor free range setting focuses on cattle. 

Papers   \cite{schwager2007robust, guo2006animal, kuankid2014classification, wietrzyk2009enabling} categorize the periods of animal activity and inactivity using accelerometer, pedometer or magnetometer to measure the position and head angle of cows. The paper uses GPS and a local server to evaluate data. Similarly, \cite{llaria2015geolocation} monitors the behavior of cattle with a collar equipped with geolocation devices and communication interfaces to determine the location of animals in mountain pastures. In   \cite{olesinski2007fuzzy}, they propose a system to tag cows with wireless devices and sensors to locate and track their movement.  Paper   \cite{wietrzyk2007energy} proposes a mobile ad-hoc network systems with routing protocols enabling low power consumption of sensors. The animals are fitted with built-in accelerometers for feed intake, and pedometers for walking intensity. 

Paper \cite{nagl2003wearable} focuses on monitoring the health of cows, collects and analyzes data obtained from sensors mounted on cattle.  The proposed system controls the sensors wirelessly with a microcontroller and uses GPS to control cows movement.  A system to detect diseases or pregnancy is presented in  \cite{wietrzyk2008practical}. The collars used a built-in accelerometer, and a pedometer to measure the intensity of feed intake. With similar focus on the health of cows, \cite{kwong2009adaptation} proposes a real-time health monitoring system, whereby the collars are equipped with antenna, relay routers, and base stations. The system focuses on low cost, low power consumption smart sensing and computing, and incorporates solar energy.   Paper \cite{harris1998ambulatory} monitors nervous system activity and cardiovascular system response in sheep. They propose a system called \emph{ Free Range Physiological Monitor} to be attached to the back of a sheep to record and process raw data for analysis. In some cases environmental monitoring is also subject of research. Paper  \cite{butler2004virtual} develops virtual fences that can control animals movement and space without man made permanent structures. The cows are equipped with a smart collar consisting of a GPS unit and a sound amplifier.

\begin{table*}[ht!]
\centering
\caption{Outdoor farm animals (cattle, except sheep in \cite{harris1998ambulatory})}
\label{tab:outdoor}
\begin{tabular}{@{}ccc@{}}
\toprule
\multicolumn{1}{c}{Application}  & \multicolumn{1}{c}{\begin{tabular}[c]{@{}c@{}}Papers\end{tabular}}  & \multicolumn{1}{c}{\begin{tabular}[c]{@{}c@{}}Technology\end{tabular}} \\ \midrule
     Behavioral Monitoring & \begin{tabular}[c]{@{}l@{}}\cite{schwager2007robust, butler2004virtual, wietrzyk2007energy, llaria2015geolocation, guo2006animal,kuankid2014classification, wietrzyk2009enabling, olesinski2007fuzzy} \end{tabular} &  \begin{tabular}[c]{@{}l@{}} GPS, servers, sensors, base station, antena collar, accelerometer, \\pedometer, ADC, modems, microprocessors \end{tabular} \\  \midrule
     Animal Health & \begin{tabular}[c]{@{}l@{}}  \cite{wietrzyk2008practical, kwong2009adaptation, harris1998ambulatory, warren2003distributed, nagl2003wearable} \end{tabular}& \begin{tabular}[c]{@{}l@{}} Sensors, processor boards, solar systems, Bluetooth, telemedicine \end{tabular} \\ \midrule  
\end{tabular}
\end{table*}


\begin{table*}[ht!]
\centering
\caption{Classification based on technologies used in farm industry}
\label{tab:farm2}
\begin{tabular}{@{}lll@{}}
\toprule
\multicolumn{1}{l}{\begin{tabular}[l]{@{}l@{}}Technology\end{tabular}} & \multicolumn{1}{l}{Applications}  & \multicolumn{1}{l}{\begin{tabular}[l]{@{}l@{}}Research work\end{tabular}} \\ \midrule      
      \begin{tabular}[l]{@{}l@{}} GPS sensors \end{tabular} & \begin{tabular}[l]{@{}l@{}} Tracking and positioning, behavior monitoring \end{tabular}& \begin{tabular}[l]{@{}l@{}} \cite{schwager2007robust},\cite{butler2004virtual}, \cite{llaria2015geolocation}, \cite{guo2006animal},\cite{wietrzyk2009enabling},\cite{nagl2003wearable}   \end{tabular}   \\ \midrule
     \begin{tabular}[l]{@{}l@{}} Accelerometer \\sensor  \end{tabular} & \begin{tabular}[l]{@{}l@{}} Measures the acceleration as a function of time, i.e. the movement\\ Examples: feed intake, renuminaration, by attaching it in the collar \end{tabular}& \begin{tabular}[l]{@{}l@{}}   \cite{wietrzyk2008practical}, \cite{wietrzyk2007energy},\cite{guo2006animal}, \cite{kuankid2014classification}, \cite{wietrzyk2009enabling},\cite{banerjee2012remote} \end{tabular}   \\ \midrule
     \begin{tabular}[l]{@{}l@{}} Pedometer \\sensor \end{tabular} & \begin{tabular}[l]{@{}l@{}}  Low cost device, usually attached to legs for activity monitoring, \\such as counting number of steps \end{tabular}& \begin{tabular}[l]{@{}l@{}} \cite{wietrzyk2008practical}, \cite{wietrzyk2007energy},\cite{guo2006animal}, \cite{wietrzyk2009enabling}   \end{tabular}   \\ \midrule
     \begin{tabular}[l]{@{}l@{}} ECG, Pulsoxymetry  \end{tabular} & \begin{tabular}[c]{@{}l@{}} Heart/respiratory rate monitoring \end{tabular}& \begin{tabular}[c]{@{}l@{}} \cite{warren2003distributed},\cite{sieber2012wireless}   \end{tabular}   \\ \midrule
     \begin{tabular}[c]{@{}l@{}} Body temperature\\sensors  \end{tabular} & \begin{tabular}[c]{@{}l@{}} Also surgically implantable (in pigs) \\ used to detect fever and body temperatureduring oestrus period (cows) \end{tabular}& \begin{tabular}[c]{@{}l@{}}   \cite{warren2003distributed},\cite{mudziwepasi2014assessment},\cite{mccauley2005wired}  \end{tabular}   \\ \midrule
     \begin{tabular}[l]{@{}l@{}}  Temperature and \\ humidity sensors \end{tabular} & \begin{tabular}[c]{@{}l@{}} Ambient sensing of the farm houses (indoor), \\ maintained to comfort the animals and also for better yield (milk, meat)  \end{tabular}& \begin{tabular}[c]{@{}l@{}}  \cite{sarangi2014development}, \cite{jindarat2015smart},\cite{ammad2014wireless},\cite{congguo2010intelligent},\cite{lee2010design},\\ \cite{zhang2012design},\cite{arvanitis2007nonlinear}, \cite{noor2013temperature},\cite{mccauley2005wired}  \end{tabular}   \\ \midrule
     \begin{tabular}[c]{@{}l@{}}  Camera \end{tabular} & \begin{tabular}[l]{@{}l@{}} health/behavior monitoring using image processing  \end{tabular}& \begin{tabular}[l]{@{}l@{}}  \cite{lee2010design},\cite{weixing2010detection},\cite{poursaberi2011online}  \end{tabular}   \\ \midrule
     \begin{tabular}[l]{@{}l@{}} ZigBee  \end{tabular} & \begin{tabular}[l]{@{}l@{}} A wireless application and network layer protocol \\
      Low cost, low power and mesh networking \end{tabular}& \begin{tabular}[l]{@{}l@{}}  \cite{zhang2012design},\cite{ma2012monitoring},\cite{li2010design}   \end{tabular}   \\ \midrule
     \begin{tabular}[l]{@{}l@{}} WSN  and MANET\end{tabular} & \begin{tabular}[l]{@{}l@{}} Wireless Sensor Network (WSN) and Mobile Adhoc network (MANET)\\ motes (including sensors) communicate wirelessly, \\ data is usually processed at the server, and \\also partially at motes  \end{tabular}& \begin{tabular}[l]{@{}l@{}}  \cite{wietrzyk2008practical},\cite{wietrzyk2007energy},\cite{kwong2009adaptation}, \cite{guo2006animal},\cite{wietrzyk2009enabling},  \cite{banerjee2012remote},\\\cite{olesinski2007fuzzy},\cite{sarangi2014development},\cite{ammad2014wireless},\cite{zhang2012design},  \cite{mccauley2005wired},\\\cite{nagl2003wearable},\cite{sieber2012wireless}, \cite{ma2012monitoring},\cite{li2010design},\cite{mudziwepasi2014assessment}  \end{tabular}   \\ \midrule
\end{tabular}
\end{table*}


\section{Wild Animals: No Rules other than Nature}\label{wild}
The dominant connectivity and sensing technology for wild life monitoring, protection and scientific research of their behavior and physical characteristics are wireless sensor networks. This technology exhibits exceptionally low battery power consumption and is designed to be ultra-light weight. These systems are robustly engineered to deal with intermittent connectivity, due to either the animal behavior or environmental factor, and endure various climate and environmental conditions. Another category of more recent work proposes a more generic IoT technology framework, as the evolution of wireless sensor networks moves towards more heterogeneity, including wireless cellular networks, alternative versions of radio technologies or even unmanned aerial vehicles (drones) \cite{DosSantos2015}. Of note is the importance of visual sensing through cameras, which due to the limited bandwidth of wireless sensor networks has not been deployed to reach their full potential, such as for visual recognition or in-situ image processing. 

\subsection{Tracking} 
Human curiosity and fascination with nature has motivated research in wild animal tracking much before the invention of the Internet. Today's technology makes it possible collect, process and visualize data, both in real time and for long term, and sustained research into animal species. Tracking devices are often designed to know the exact position of the animal, and track the motion of the animal through an acceleration sensor, with a low power consumption \cite{Song2011}. In general, the design of the smart system can be inspired by specific animal behavior, leading to new innovation in network architecture, or existing technologies and systems can be adapted to the animal behavior leading to innovation in animal welfare applications and scientific discovery. Finally, just like domestic and farm animals, wild animals can also be tracked with wearable or non-wearable sensor systems.

A  novel hybrid architecture \cite{Anthony2012} for monitoring whooping cranes, an endangered species, uses global infrastructure (cellular networks) during their annual migration of 4,000 km, and a short range, ad-hoc networks in breeding and nesting grounds. This platform led to a new class of so-called \emph{cellular sensor networks}. The focus in \cite{Huang2010} is on a low cost high-sparse WSN system prototype for tracking of multiple species in the same environment. Paper \cite{Zviedris2010} proposes LynxNet, a monitoring GPS based system using collars on lynx in delay tolerant networks. An early work on wild animal tracking presented in \cite{Juang2002}, also known under the name Zebranet, proposes a wireless peer-to-peer (ad-hoc) sensor system, minimizing energy consumption and storage, to support wildlife tracking for biology research across large geographic areas. One of the interesting challenges addressed in a system called Virtual-Beacon \cite{Radoi2015} used to track wild horses are methods for uploading sensor data from mobile nodes to base stations in nodes with limited power. Mountain lions were studied in \cite{Rutishauser2011} with a network system composed by mobile and static nodes with sensing, processing, storage and communication. A generic wild animal monitoring system framework \cite{Picco2015} uses geo-referenced proximity detection, with an adaptive model for multiple species suitable for biologists to conduct the research on various species. Monitoring techniques based on ZigBee and GPRS \cite{Tan2011} have been used on two species of protected monkeys under risk in the Mexican jungle. Simulations of hotspot-based WSN routing algorithms \cite{Anand2013} have been used to monitoring of protected wild tigers.

The research reviewed so far included a specialized tracking device, or a GPS based collar. Tracking based on sensors located in corresponding geographic areas with in-situ animal detection systems is also possible. One of the pioneering work \cite{Mainwaring2002} in this space, known as DuckIsland, designs and develops a complete WSN for habitat monitoring. The aim of this work is to understand the behavior of wild animals, especially in islands where the presence of humans can disturb the breeding. Other work \cite{He2016,Bagree2010, Liu2011} focus on integrated camera-sensor network systems, image processing for animal detection, tracking, species classification and cloud based data management that includes a web interface. Other approaches use ultra-low-power sensor systems (and low weight $<2$gr) \cite{Dressler2016} for tracking bats, and a WSN system based on grid positioning \cite{Joshi2008} for tracking turtles. Paper \cite{Tapiador-Morales2015} presents a new monitoring system for wild animals using inertial sensors, which transmits the information using ZigBee technology. A group of papers \cite{Dyo2009, Dyo2010, Dyo2012} focus on automated and sustainable wildlife monitoring systems with RFID for badgers. 

There are solutions that do not solely focus on wireless network system technologies, but also on visualization and processing of data collected, in form of various frameworks. A framework for processing, analysis and visualization of tracking data for wild animals \cite{Hunter2013} has been used to study animal behavior and ecology in Australia. Web frameworks (website, databases) \cite{Currier2015, Constantinescu2013} have been used for collaborative aquatic animal tracking. A wildlife monitoring and communication system \cite{Liu2015} was a part of an innovative proposal for the tracking and recognition of wild animals. An integrated video and sensor system \cite{He2008} was mounted directly on the animal to record all what deers see. An animal-to-animal Internet sharing capability method is proposed in \cite{Nakagawa2014} in order to maximize monitoring performance in inaccessible areas. The authors in \cite{Peng2012} propose a monitoring system to detect wild animal and poachers in natural protection zones and alert users sending pictures or videos. Finally, paper \cite{Dyo2010} proposes the use of magnets for the localization of underground animals with the help of receiver antennas, which ensures monitoring over long periods of time. The magneto-inductive tracking can be used for any type of underground animal species.

\subsection{Human-animal cohabitation} 

Human-animal cohabitation concerns welfare of wild and feral animals in the areas populated by humans, both in rural and urban areas.  A smart system \cite{Tennakoon2015} in Sri Lanka was to detect breakages in fences used to keep wild elephants away from humans for the protection of both humans and animals. Similarly, a WSN system that uses passive nodes and infrasonic sounds \cite{Mathur2014} was deployed in India to deter elephants from crossing railways. The use of IR sensors and seismic sensors was proposed in \cite{Nakandala2014} for detection of wild elephants entering villages. A new WSN system \cite{Viani2011, Viani2014} was used to alert drivers about dangerous situations caused by wildlife crossing, through the adaptive actuation of light signals. In an effort to protect sea turtle hatchlings from tourists, this research \cite{Zimmerman2014} integrates a low cost movement sensor system with wireless cellular network. A new approach \cite{Zhang2011} proposes an IoT based autonomous water conservancy system based on the actual water levels and local density of deer, an endangered species in China. Another approach \cite{Ilcev2014} proposes a new positioning system based on ultrasonic signals for landing flight objects used for wildlife protection.

Image processing is often used in combination with camera-based sensors for animal recognition and their protection \cite{Duran2007, Tovar2010}.  A new methodology \cite{Diaz2012} applies compressive sensing for sound recognition and classification in WSN systems in order to minimize the number of samples required to reduce power consumption which is a critical issue in WSN. An IR video processing algorithm for the recognition of migratory birds \cite{Wei2014} is used to determine the optimum allocation of wind farm areas to avoid the collision with birds. Other efforts detects wild kangaroos with cameras \cite{Zhang2015}, wild animals in snow \cite{Oishi2010}, and hidden fawns in meadows using compression-based algorithm, radars, thermal and RGB cameras \cite{Cerra2009}. A new technique called Sparsogram is proposed in \cite{sparsogram2014} for the classification of collected audios in order to detect unlawful human intrusion in protected wild areas. A radar system \cite{Fackelmeier2009} is used to detect covered microwave reflecting objects with high quantity of water to protect fawns from death during the pasture mowing.

\subsection{Wild animals in confinement} 

Wild animals in confinement, including zoos and animal sanctuaries have been subject to research by the ACI science community, with the purpose of animal welfare but also to study the human-animal interactions for education and conservation. Paper \cite{Carter:2015:NAA:2832932.2837011} discusses the role and opportunities that ACI and new technologies can play in zoos to improve the animal welfare. In \cite{Pons2014}, a few scenarios for the so-called Intelligent Playful Environment for Animals (IPE4A) are proposed to help animals overcome isolation, poor physical condition, repetitive training exercises, or remote digital interaction with humans. 


\begin{table*}[ht!]
\centering
\caption{Wild animals: applications, devices, and systems}
\label{tab:wild}
\begin{tabular}{@{}cccc@{}}
\toprule
\multicolumn{1}{c}{Application}  & \multicolumn{1}{c}{\begin{tabular}[c]{@{}c@{}}Papers\end{tabular}}  & \multicolumn{1}{c}{\begin{tabular}[c]{@{}c@{}}Devices\end{tabular}} & \multicolumn{1}{c}{\begin{tabular}[c]{@{}c@{}}Systems\end{tabular}} \\ \midrule
     Tracking & \begin{tabular}[c]{@{}l@{}}\cite{Song2011, Anthony2012, Huang2010, Zviedris2010, Juang2002, Radoi2015, Rutishauser2011, Picco2015, Tan2011,Anand2013, He2016,Dressler2016,Bagree2010,Joshi2008,Tapiador-Morales2015, Liu2011, Dyo2009, Dyo2010, Dyo2012, Mainwaring2002, Hunter2013, Currier2015, Constantinescu2013, Liu2015, He2008, Nakagawa2014, Peng2012} \end{tabular} &  \begin{tabular}[c]{@{}l@{}} GPS collars, wearable tracking devices, camera traps\\ miniature sensors, Xbee sensors, IR sensors\\ RFID sensors, magnetic sensors \end{tabular} &  \begin{tabular}[c]{@{}l@{}} Hybrid cellular/ad-hoc networks\\Highly sparse WSN\\ZigBee and GSM/GPRS\\ Delay Tolerant Networks (DTN) \\ Camera and proximity detection systems\\Grid positioning \\ IR image sensor network\\Magnetic localization \end{tabular}\\  \midrule
     Co-habitation & \begin{tabular}[c]{@{}l@{}}  \cite{Tennakoon2015, Mathur2014, Viani2011, Viani2014, Zimmerman2014,Zhang2011,Ilcev2014,Duran2007, Tovar2010,Diaz2012, Wei2014, Zhang2015, Cerra2009, Oishi2010, sparsogram2014, Fackelmeier2009, Nakandala2014} \end{tabular}& \begin{tabular}[c]{@{}l@{}} Fencing sensors,  IR sensors, Seismic sensors\\Doppler radars, antennas, IR cameras\\Camera sensors, thermal and RGB cameras\\Mobile phone \end{tabular} &  \begin{tabular}[c]{@{}l@{}} Electric fences, Virtual fences \\
Infrasonic sound systems\\
Lights signals for drivers\\
GSM, GPRS, 3G, WiFi, Zigbee\\
WSN with image processing\\
WSN with habitat monitoring \\
IR video systems\\
Deformable Part Model detection systems \\
Sparsogram\\
Microwave systems\\
 \end{tabular}\\ \midrule 
     In confinement & \begin{tabular}[c]{@{}l@{}}  \cite{Carter:2015:NAA:2832932.2837011,Pons2014} \end{tabular}& \begin{tabular}[c]{@{}l@{}} Mobile devices, servers\\
Sensors, cameras
 \end{tabular} &  \begin{tabular}[c]{@{}l@{}} Wireless networks (cellular, WiFi, sensor)\\
Augmented Reality\\
     Intelligent Playful Environment for Animals (IPE4A)
 \end{tabular}\\ \midrule   
\end{tabular}
\end{table*}



\section{Findings and Opportunities}
\label{findings}
This section summarizes the main findings from the review over various categories of the smart systems, including sensors, networks and computing systems. We discuss economic factors that drive all aspects of systems engineering and design, and their impact on animal welfare. We conclude this section with some research opportunities, many that are applicable in general to all research in smart systems but some that are unique to animal welfare.

\subsection{Wearable vs. non-wearable sensor systems}
For domestic and wild animals, wearable sensors have been primarily used for GPS-based tracking. The emphasis is on engineering compact and light-weight designs to minimize animal discomfort, and improve reliability since animals can destroy devices that make them uncomfortable. Depending on the species, wearable trackers can be attached either to collars (dogs) or legs (birds). Working domestic animals, like service dogs, or search and rescue dogs,  are mostly equipped with wearable jackets with multiple sensors that depend on the application. Jackets are considered a better choice over collars from an animal welfare perspective since they distribute the weight of the wearable system on animal body. In some cases, dog's vests include a combination of multi-purpose sensors, combining health, tracking and human-health related sensors, such as with vibrotactile feedback. For farm animals, on the other hand, the driving factor is the cost and the accuracy of sensor data gathered, so the weight of wearables plays less of a role. For instance, collars used in cattle farming can include an active transponder, antenna, accelerometer and GPS sensors. Some temperature sensors are also surgically implantable as in the case of pigs. Alternatively non-wearable system for monitoring and tracking appear in form of kinetic sensors, or ambient sensors of the building where animals are held. Ambient monitoring is especially important in indoor farming, where it is used to measure temperature and humidity. Non-wearable systems are in comparison less developed overall, and carry potential for improving animal welfare through further innovation. While most of the sensor systems reported support network connectivity, only some of the sensor systems reviewed connect to common shared infrastructure like the cloud. This makes it hard to do longitudinal tracking, and share data and best practices.

\subsection{Networked remote sensing}
The diversity of animal species reflects the richness and heterogeneity of wireless technologies used for animal tracking. Most notably in the domain of wild animal tracking, multiple types of hybrid wireless networks were reviewed. These range from integrated cellular and ad-hoc networks, to wireless sensor networks and delay tolerant networks. In many cases, the wireless network architecture needs to be adapted to the species' migratory patterns, and be designed as ultra-low-power and low cost sensor network. Of note is the integration of video and camera based wireless sensor networks, which present technical challenges with respect to bandwidth and capacity management. For networks built and operated underground, underwater and under challenging climate and geographic conditions, special attention needs to be paid to the robustness of the system. A few research papers reviewed also pointed to the issue of maintenance required for the systems built in remote wilderness settings. For domestic animals and pets, a standard-based integration with wireless 3G networks and smart phone based applications is a common approach, though there is comparably less work to address the ongoing efforts in the next generation of cellular network applications (5G). In the area of livestock agriculture in general, there are already commercial smart farm management services that help farmers to track whether livestock have enough food, water, and fresh air, while also monitoring the temperature and ensuring that they are safe and secure.  The key challenges are trade-offs between cost, battery power and network connectivity in practical dynamic monitoring scenarios. These systems could be easily augmented to provide remote veterinary care, which could prevent unnecessary loss of life, and provide further economic benefit. Robots could be employed to care for and exercise animals remotely. 

\subsection{Cloud-based applications and data processing}

The use of remote sensing science that requires network connectivity and remote access has been very effective in supporting conservation of species, habitats, communities, and ecosystems. There are GPS systems for tracking lost pets, and smart pet doors that allow one to program which pets can go in and out, and when.  The use of remote sensing in the farm setting is currently focused on providing farm operators with precision maps, crop scouting capabilities, information to aid in crop care, and more. 
Research on managing and representing the large amounts of data, including those collecting about animal sounds, pictures and videos, generated by wearable and non-wearable animal sensor network systems is in infancy. Most of the work surveyed, in all domains of animal welfare, assume a virtual connection to either a stand-alone computational node, or a distant server, but only very few extend the data processing and sharing to the cloud. Although not considered in this survey, domestic pets can currently be tracked and monitored through cloud based applications over commercial wireless cellular network in urban settings. But little is known about the data collected, and whether it can be shared, or used to track the health of animals or their living conditions.  For farm animals, where economic factors play a role related to the quality and quantity of the animal based products, web-service applications have been proposed to monitor animal growth, health and food intake. The same questions regarding the data collected exist. An example of the power of sharing this data is the United Nations Food and Agriculture Organization (FAO) pilot program in Africa\footnote{Aga in Action, Public release of the new EMPRES-i, http://www.fao.org/, last updated in September 2012.} which uses mobile phone applications to track animal vaccination and treatment campaigns, and stores this information in a global database which can be used to pinpoint and contain animal disease outbreak. 

\subsection{Economic factors}
Economic factors are perhaps most important consideration in this space. In particular, in livestock agriculture, low cost and low power devices are imperative. Cost is a real factor for even simple issues like having an unique ID for every animal. The National Livestock Identification System (NLIS) of Australia regulations require that all cattle be fitted with a RFID device in the form of an ear tag or rumen bolus before movement from the property and that the movement be reported to the NLIS. A similar system is used for cattle in the European Union (EU), each animal having a passport document and tag in each ear carrying the same number. The U.S. National Animal Identification System, which has been in development by the USDA since 2002, also promotes microchips or other ID tags to livestock so they could be monitored throughout their lifetimes by a centralized computer network. While the upfront costs of microchips, and supporting infrastructure can be expensive, these systems could be invaluable in tracking a specific animal through the entire supply chain, and quickly identifying and containing disease outbreaks. As the price of these systems continues to fall, the enormous costs incurred in such situations that end up in large scale recalls could very well make the large scale deployment of these technologies very cost effective.

\subsection{Research Opportunities}
In order to make smart technologies usable and economically viable for all three groups studied, there is a need for computer science  and engineering research in ultra-low power and ultra low-cost hardware devices, more efficient algorithms for collecting and storing large amounts of data, advances in networking, and common infrastructure and repositories to enable sharing of information, alerts and best practices in real time. To interpret and analyze the rich multi-modal data collected, there is a need for sophisticated data analytics including ones based on machine learning and natural language processing. We believe that the integration of animal welfare requirements into early design of smart cities and communities is a tremendous opportunity for reuse and sustainability. 

Inter-disciplinary work involving computer scientists, engineers, animal behaviorists, conservationists and veterinarians could yield real innovations and technologies in this space. These include novel gesture and posture recognition algorithms for all animal groups considered, and enhanced two-way communication with animals through implantable, wearable and non-wearable devices. It would be valuable to generalize the many specific instances of animals detecting natural disasters and diseases, and provide more sophisticated and interactive ways to warn and keep wild animals away from danger via wearable devices. We believe inter-disciplinary animal centric computing research is key for usable, sustainable and economical smart technologies for ensuring the welfare of all animal.

\section{Conclusion}
\label{conclusion}

This paper systematically reviewed smart technologies used in animal welfare, in three main categories of animals: domestic, farm and wild animals. A smart system, as we define it, assumes sensing and computing capabilities that are interconnected, not only with various networking technologies but also computing systems that can collect, process, and evaluate data related to the animal welfare. We defined animal welfare in generic terms, recognizing that systems reviewed serve to help animals stay healthy, free of pain and suffering, and also being positively stimulated in their environment. Many of the technologies reviewed have been used to great benefit in specific cases and situations, but having all these technologies available and integrated with a centralized database to track and share this information and best practices would have enormous societal and economic benefits.

\par The recommendations for further research include a few salient features of the systems reviewed and their potential to improving animal welfare, i.e., 
\begin{itemize}
\item \textbf{Develop integrated and open cloud based systems, applications and services}. Even though research has been reported on smart farming and agriculture, much work is to be done in integrating the specialized sensor network system with the current cloud services and infrastructure and opening the data and systems for sharing, programmability and further innovation. 
\item \textbf{Integrate cross-species and cross-sectorial research}. We have found a lot of common features in how the animal based sensor network systems are built and used, but little or no evidence that the systems can be reused across species or animal applications. For instance, farming system can much benefit from the knowledge in low cost, and low power wild animal tracking, as well as from wearable systems for dogs. 
\item \textbf{Include animal centered research in smart agriculture}. Even though the smart agriculture concepts do not exclude animals, much of the focus today is on plant-based agriculture, and comparably less on livestock agriculture. 
\item \textbf{Integrate topics of animal welfare conceptually into smart "X" systems and the IoT world}. Smart and connected cities and communities are now becoming a reality. This is a perfect opportunity to add animal welfare to the agenda. For little or no extra cost, these technologies can be also be used to track bird and other wildlife migration pattern, track and find missing pets and livestock, predict natural disasters, and a host of other possible applications. Smart transportation can be used to monitor the welfare of transported animals, smart energy can be used to track animals outdoors, smart cities can monitor wild animals in cities, and domestic animal applications can be integrated in smart homes.  
\item \textbf{Create smart emergency and disaster response for animal welfare}. All animals, be it pets, farm, zoo or wildlife, are arguably the biggest casualties in emergencies and disasters like fires, earthquakes, floods and other natural disasters. In such situations, when first responders are stretched to the limit, smart technologies can play a significant role from detection to prevention to recovery. Smart systems can detect the emergency, the number and kinds of animals in need, and take predetermined rescue and recovery measures. 
\item \textbf{Make animal welfare economically sustainable}. As this review shows, animal welfare can be economically sustainable, when supported through low cost smart systems, or when integrated into systems already in place. The data provided by technologies can inform consumers of animal products of the provenance of the livestock, and provide strong economic incentive and aid adoption. 
\item \textbf{Use smart technologies to learn from the animal world}. As part of the ACI, there are untapped opportunities to use smart technologies learn from the animal world. There is documented evidence that animals can provide early warnings for impending natural disasters like earthquakes, floods and hurricanes, and diseases like heart attacks, cancer or diverse types of seizures. But smart technologies present the possibility to scale this from isolated and often unrelated cases into an actionable methodology that could have enormous benefits. 
\item \textbf{Promote Education and Awareness}. The key challenge in adoption of any of these smart technologies is lack of awareness of the existence, effectiveness and economic benefits within the farming community, among consumers, and even technologists. Educating the veterinary and wildlife conservation communities about smart technologies could also make great strides in increasing deployment. Computer science and engineering curricula need to include syllabi on smart technologies and systems for animal welfare. 
\end{itemize}
There are undoubtedly hard technical and economic challenges to overcome, but these are minor in comparison to changing the existing mindset. As this review demonstrates, there are many smart technologies in use today, and a sea of promising innovations in the future, making it possible for smart computing and sensing technology to co-exist with the animals in a sustainable, humane and mutually beneficial manner.

\section*{Acknowledgment} The authors would like to thank (alphabetically) Marcel Caria, Francisco Carpio, Ashwin Gumaste, Sandeep Singh, and Birgit U. Stetina for their advice, help and support. This work is dedicated to the memory of Dr. Rujuta Kitty Raichura.

\section*{Disclaimer} Any opinion, findings and conclusions or recommendations expressed here do not necessarily reflect the views of the National Science Foundation (NSF).

\bibliographystyle{IEEEtran}
\bibliography{Animal}

\end{document}